\documentclass[conference]{IEEEtran}
\usepackage{cite}
\ifCLASSINFOpdf
\usepackage[pdftex]{graphicx}
\else
\usepackage[dvips]{graphicx}
\fi
\usepackage{array}
\usepackage{color}
\usepackage{amsmath}
\usepackage{amssymb}
\usepackage{amsthm}
\usepackage{titlesec}
\usepackage{graphicx,epstopdf}
\usepackage{tabularx}
\usepackage{comment}

\usepackage[none]{hyphenat}

\usepackage{array}
\usepackage{multirow}
\titlespacing*{\section}{0pt}{1.0ex}{0.5ex}    
\titlespacing*{\subsection}{0pt}{1.2ex}{0.1ex}

\makeatletter

\begin{document}
\title{Cyber-Physical  Power  System  Layers:  Classification,  Characterization, and  Interactions}
\author{Michael Abdelmalak, \emph{Member, IEEE}, Narayan Bhusal, \emph{Member, IEEE}, \\ Mukesh Gautam, \emph{Graduate Student Member, IEEE}, and Mohammed~Benidris, \emph{Senior Member, IEEE}\\ Department of Electrical and Biomedical Engineering, University of Nevada-Reno, Reno, NV 89557, USA \\ Emails: \{mabdelmalak, bhusalnarayan62, mukesh.gautam\}@nevada.unr.edu, and mbenidris@unr.edu\vspace{-1ex}} 


\maketitle

\begin{abstract}
This paper provides a strategy to identify layers and sub-layers of cyber-physical power systems (CPPS) and characterize their inter- and intra-actions. The physical layer usually consists of the power grid and protection devices whereas the cyber layer consists of communication, and computation and control components.   
Combining components of the cyber layer in one layer complicates the process of modeling intra-actions because each component has different failure modes. On the other hand, dividing the cyber layers into a large number of sub-layers may unnecessarily increase the number of system states and increase the computational burden. In this paper, we classify system layers based on their common, coupled, and shared functions. Also, interactions between the classified layers are identified, characterized, and clustered based on their impact on the system. Furthermore, based on the overall function of each layer and types of its components, intra-actions within layers are characterized. The strategies developed in this paper for comprehensive classification of system layers and characterization of their inter- and intra-actions contribute toward the goal of accurate and detailed modeling of state transition and failure and attack propagation in CPPS, which can be used for various reliability assessment studies.

\end{abstract}

\IEEEpeerreviewmaketitle

\section{Introduction}
The advancements in communication and automation technologies have increased significantly in the last decade resulting in widespread integration and deployment in power systems. Grid modernization approaches created what is now known to be cyber-physical power systems (CPPSs). Such systems are composed mainly of cyber layer, i.e., communication and control systems, and physical layer referring to the power system. Despite the noticeable added benefits of cyber layer on power systems achieving reliable, secured, and economic operation, increased vulnerabilities against cyber threats and attacks are always being associated with the level of integration. Also, the reliance to leverage user-friendly human interface platforms, cloud computation, and smart artificial-intelligence devices create further complexities to analyses of CPPSs. Therefore, it has become a necessity to accurately model the state transitions and propagation behaviors in CPPSs for improved evaluation and enhancement of their resilience and performance.  

Recently published research in \cite{9763485, 9167203, graja2020comprehensive}, provides a comprehensive review of CPPSs from the perspective of modeling, simulation, and analysis with cyber security applications. This paper also provides literature survey on cyber attacks and cybersecurity measures for CPPSs. This work describes the CPPS as the coupled network of cyber and physical systems. Cyber layer consists of computation, communication, and control systems. Physical system, on the other hand, consists of a physical power grid governed by physics-based rules. In \cite{carreras2020conceptualizing}, key features of cyber-physical systems in multi-layered architecture are conceptualized. This work characterizes the cyber physical system into physical layer, cyber-physical layer, and the cyber layer. Physical layer consists of physical components and their dynamics, physical measurements, and physical operators. Cyber-physical layer includes programmable controllers, real-time communication networks, sensors, and actuators. Cyber layer is formed by a combination of cyber communication networks, supervisory computers, and supervisors. 

Cyber layer can be identified as the layer responsible for the computation, analysis, and assessment of the power system on the regional and global scale. Defining the boundaries of a cyber layer within a CPPS model is not a straightforward process. First, the advancements in information and communication technology have resulted in embedded smart computation processors in all power system components. This raises a concern whether such computation parts are system or component involved. Also, some system computational tasks take place at the local level such as protection decisions; whereas other wide-area analyses are handled in the energy management systems \cite{7936473}. This raises a concern whether the cyber layer is composed of a single layer or can be split into several layers. The cyber layer comprises all required applications to maintain reliable and economical operation of the power system. Some of these applications are run in the local level prior to passing to global level such as automatic generation control, remedial action schemes, and protection protocols. Other global applications include but are not limited to state estimation, real-time contingency analysis, security constrained optimal power flow, unit commitment, and energy market optimization. Determining the proper input data into diverse applications causes a confusion on boundaries of the cyber layer. 

Whereas the power grid represents only a physics-governed physical layer, the cyber layer consists of several layers such as sensor, protection, communication, computation, and control layers. Combining the components of the cyber layers in one layer complicates the process of modeling intra-actions because each component has different failure modes. On the other hand, dividing the cyber layers into a large number of sub-layers may unnecessarily increase the number of system states and increase the computational burden. Therefore, rigorously identifying system heterogeneous layers (cyber and physical) and comprehensively characterizing their inter- and intra-actions are essential to (1) establish accurate models for state transitions; (2) identify chains of failure propagation within and between layers; and (3) develop efficient and practical reliability and resilience analysis, evaluation, and enhancement methods and strategies for CPPSs. Further research is inevitable for the maturity of CPPS classification, characterization, and modeling, simulation, and analysis of interactions between and within the CPPS layers.

This paper establishes strategies to identify CPPS layers and sublayers and characterizes their inter- and intra-actions. In this paper, CPPS layers are classified based on their common, coupled, and shared functions. During classification, we start with common intended functions, of which there are many, each of which aggregates several system components. Then, we identify coupling layers (i.e., failure of coupling layers separates two or more layers) such as the communication layer, which couples the heterogeneous physical layer and remaining layers. Next, we identify shared layers such as the sensors' layer---a shared layer between the communication and protection layers. Also, interactions between the classified layers are identified and characterized; possible interactions are discussed and clustered based on their impacts on the system. Furthermore, intra-actions within each layer are characterized based on the overall function of the layer and types of its components. The strategies developed in this paper for comprehensive classification of system layers and characterization of their inter- and intra-actions contribute toward the goal of accurate and detailed modeling of state transitions and failure and attack propagation in CPPS. This is a necessary step toward developing analysis, evaluation, and enhancement methods for CPPS reliability and resilience.

The rest of the paper is structured as follows. Section \ref{modeling CPPS} provides a survey on existing approaches of classification, characterization, and interaction of cyber-physical layers, and criteria of CPPS modeling. Section \ref{model} describes the suggested classification, characterization, and interactions between CPPS layers. Section \ref{conclusion} provides the concluding remarks.

\section{Modeling of Cyber-Physical Power Systems} \label{modeling CPPS}
Modeling of cyber-physical systems across various domains has gained significant interest in the last decade. This includes, but not limited to, biomedical systems, transportation systems, and energy systems \cite{CPSbook2020, 10.1007/978-3-540-87785-1_5}. Proper models of CPPSs are necessary for accurate, reliable, and efficient analysis and assessment \cite{sztipanovits1997model, 9893118, yohanandhan2022specialized}. This section summarizes the most recent modeling approaches of CPPSs and the associated dependencies across the model layers. Also, it presents few criteria to measure the capabilities of these models within the Cyber-Physical domain.      

\subsection{Existing CPPS Models}

The layer classification of the CPPS model varies in the existing literature based on the study or the system. In \cite{IET1697132, 7884993},  a two-layer CPPS model has been provided to assess the transient power system stability against control and communication failures. The first layer represents the power grid system, whereas the second refers to the cyber layer. Another two-layer CPPS model has been provided in \cite{MPCE2018944957}, where the cyber layer is represented by three sub-layers including measurements, protection, and control. Authors of \cite{7936473} have restructured the CPPS model in \cite{MPCE2018944957} to include an intermediate layer between the cyber and physical layers. The connecting layer handles three main applications, wide-area monitoring, protection, and control. The function of the intermediate layer has been changed in \cite{6473865} to represent only the communication between the physical layer and the cyber layer. A comprehensive four layers CPPS model has been provided in \cite{vellaithurai2015real} representing physical, communication, control, and monitoring layers.  

\begin{figure}[htbp]
\centering
\includegraphics[scale=1]{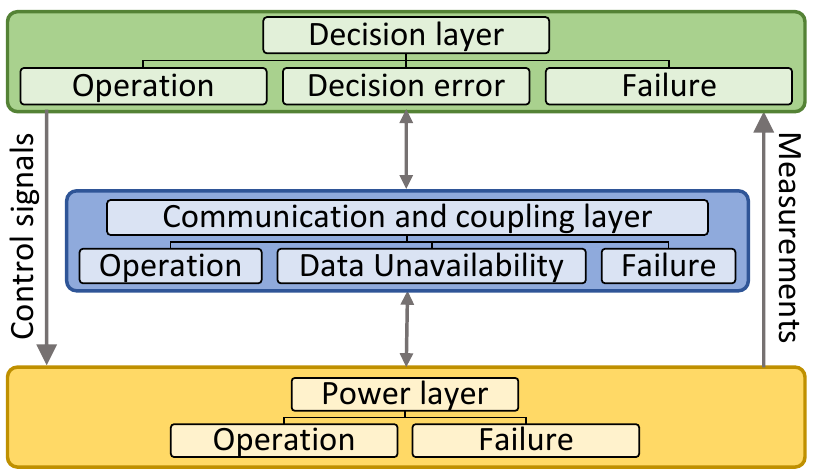}
\caption{CPPS model in \cite{8440331}}
\label{fig:CPPS_3layers}
\end{figure}

Fig.~\ref{fig:CPPS_3layers} represents a three-layer CPPS model in \cite{8440331}. The bottom layer represents the physical power system; the intermediate layer refers to the coupling communication layer; and the top layer is the decision control layer. The measurement layer is assumed fully reliable, whereas the protection layer is ignored. The mathematical computations are integrated within the control layer. It is worth noting that this model captures only the states and interactions of three main layers neglecting the inter-actions within each layer.

\begin{figure}
\centering
    \includegraphics[scale=0.99]{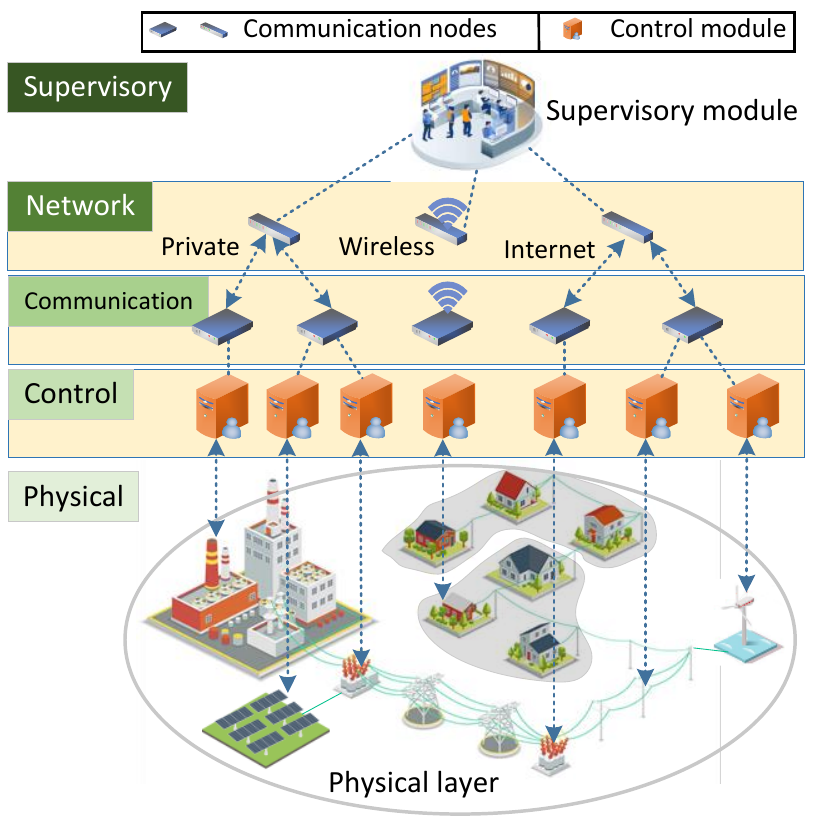}
    \caption{Different CPPS layers with the control system \cite{6016081} \cite{Zhu2019}.}
    \label{fig:zhu}
\end{figure}

A more detailed CPPS model has been developed in \cite{6016081, Zhu2019}, as shown in Fig. \ref{fig:zhu}. The model splits the cyber-physical smart grid into a hierarchical six layers including management layer, supervisory layer, network layer, communication layer, control layer, and physical layer. The presented model complies relatively with the NIST smart grid conceptual model \cite{gopstein2021nist}. The control layer includes sensors, actuators, and intrusion-detection devices. The communication layer is the connection medium between the control layer and various network types. The data routing and network formation are handled in the network layer. The computational data analysis, performed in the supervisory layer, is passed to the management layer for proper decision making. Also, the management layer takes into account the energy market, regulatory policies, and system operation.

\subsection{Dependencies in CPPS Layers}

Several studies have been conducted to model dependencies among CPPS layers \cite{yang2021reliability}. Graph theory, complex-network methods, finite state models, Petri net models, correlation methods, and cellular automate methods are some methods to model such dependencies \cite{9167203}. Five mathematical models have been presented in \cite{ tondel2018} to analyze interdependencies of CPPS layers including dynamic analysis, topological analysis, consequence analysis, causal analysis, and hazard identification. A graphical network model has been integrated with a chaotic levy flight algorithm to assess the transition of a cyber-attack to a cascading failure scenario of power grids. To model the transition between power and cyber layers on the component level, a Markov state model has been presented in \cite{6666922}. Authors of \cite{laprie2007modelling} have provided a Petri net model to capture the interdependencies between information layer and physical layer against malicious attacks. A correlation matrix approach has been introduced in \cite{7478112} to study the propagation behavior of cyber-induced failures into power systems. The cyber-physical interface matrix can be calculated using the IEEE-61850 communication scheme and available failure rate of cyber-related components.  

Various methods have been presented to classify dependencies in CPPS models. In \cite{6175593}, a classification based on the relationship between network and system elements has been introduced including both direct/indirect element-element and element-network models. Three levels of interactions have been introduced in \cite{9167203} including computational-communication interactions, communication-physical interactions, and local physical-controller-protection interactions. A comprehensive guideline to model interactions between power system layer and ICT layers has been introduced in \cite{tondel2018}. Such interactions are; (1) common cause, where the cause of failure in both systems is the same, e.g., whole substation shutdown, (2) cascading cause, where a failure in one layer propagates to another layer, e.g., power outage of communication systems, and (3) escalating cause, where an existing failure in one layer worsens an independent failure in another layer, e.g., failure in protection layer during a faulted power system. Authors of \cite{laprie2007modelling} have classified interdependencies between infrastructure layers into type of interdependencies, infrastructure environment, couplings among layers, infrastructure characteristics, state of operation, and type of failure.

\subsection{Modeling Criteria}
Though extensive research has been conducted in modeling CPPSs, only a few papers have given interest to evaluate the developed models. Selecting a particular model is a sophisticated process that requires highlighting the pros and cons of each model. Also, the compatibility of a CPPS model to a specific study or application plays a vital role in the decision process. A few main criteria are used to quantify CPPSs models including: (1) accuracy, (2) scalability, (3) fidelity, (4) application-compatibility, (5) dynamics-adaptability, and (6) topological-suitability. These metrics are explained as follows.

\noindent(1) \underline{Accuracy:} Modeling accuracy refers to the capability of a model to reproduce experimental data that agrees with the physical phenomena precisely.  In other words, this criterion measures the consistency of a model against varying scenarios and diverse input data. It is a necessity for CPPS models to maintain consistent outcomes under various constraints and diverse factors such as geographic locations and operating conditions. 

\noindent(2) \underline{Scalability:} The scalability feature refers to the capability of a model to adapt to large-scale systems and provide comprehensive representation of the system. Building a scalable CPPS model requires extensive caution with sophisticated conversion procedure, available computational capabilities, different modeling domain, diverse interoperability issues, and fast market technology.

\noindent(3) \underline{Fidelity:} If the model outcomes match the results of real-world systems, then a CPPS model is said to maintain fidelity. In CPPS, high nonlinearity levels in the power system layer impose further complexities to achieve fidelity. Due to modeling approximations, a small discrepancy can be noticed between the CPPS model and the real-world system. Maintaining least discrepancies yields high fidelity models.  

\noindent(4) \underline{Application-compatibility:} The level of information and approximation of a particular model may change based on the application or problem under study. For instance, reliability-based studies of power systems do not usually require dynamic system information. A CPPS model is said to maintain a high level of application-compatibility if it can be used across different types of studies with minimal modifications.  

\noindent(5) \underline{Dynamics-adaptability:} Power systems are characterized by high dynamics level. In various studies, it is required to capture the small-time variations in the system dynamics. This criterion aims to quantify the capability of a CPPS model to capture the dynamical behavior, particularly transient and subtransient changes in the power system.  

\noindent(6) \underline{Topological-suitability:} The NIST smart grid conceptual model describes future CPPSs in terms of seven main domains including customer, distribution, transmission, generation, market, service providers, and business services. A CPPSs model shall be capable of representing these domains, their distinctive features, and their dependencies. Due to the large-scale integration of distributed energy resources  and increased number of local control centers, the system topology is changing from a centralized structure to a distributed structure. The topological-suitability criterion shows the degree of a CPPS model to represent the new meshed distributed system topology. 

\section{Suggested Model for CPPS}\label{model}
CPPS is the combination of various layers that interact together for a reliable operation of the power grids. The power grid is usually represented as a physical layer, whereas the cyber layer might consist of several layers such as measurement, protection, communication, computation, and control layers. Combining various components of the cyber layers in one layer results in improper modeling of dependencies among components and layers. Also, it complicates the process of modeling intra-actions because each component has different failure modes. On the other hand, dividing the cyber layers into numerous sub-layers may increase the computational complexity due to the large number of system states. Therefore, accurately classifying system layers such that the inter- and intra-actions between and within them while reducing the modeling complexity and computation burden has become important.     

\begin{figure}[htbp]
\centering
\includegraphics[scale=1.01]{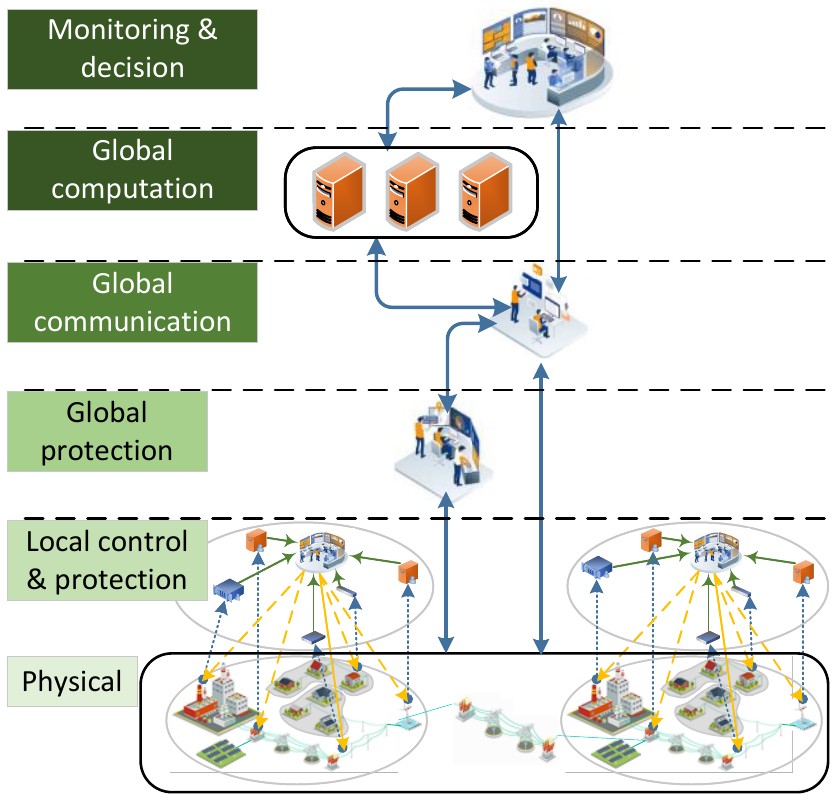}
\caption{Proposed CPPS layers.}
\label{fig:proposed_CPES}
\end{figure}

By taking the trade-off between the modeling accuracy and computational complexities into consideration, a five-layer CPPS model is identified. These layers are classified based on their common, coupled, and shared functions. The main layers are the physical grid, the global protection layer, the global communication layer, the computation layer, and the monitoring and decision layer as shown in Fig.~\ref{fig:proposed_CPES}. This architecture also consists of some local layers, for example, local protection, control, and communication layers that are not directly connected to the main monitoring and decision layer. Brief description of these layers is provided as follows. 

\subsection{Physical Power Grid Layer}
Conventional power grid is the main building block upon which the concept of CPPS has advanced. This layer provides the detailed description of the power system model, its configuration, electrical characteristics, and topology \cite{7103368}. This layer might include devices such as measurement devices and protection devices that are directly connected to power system components for proper operation and functioning of the system\cite{8440331}. Each component in the physical layer has unique fundamental functions and electrical characteristics. The physical layer can be further sub-categorized based on type of components into power system components, protection components, and measurements components.

\subsubsection{Power System Components}
This part of CPPS describes the topology of power systems using single line diagrams. The power grid is categorized based on functionality into three main categories: generation, transmission, and distribution. In normal operation, generation level should be sufficient to supply load demands under consideration of all system operating constraints. 

\subsubsection{Protection Devices}
The protection layer consists of all the protective devices that either prevent or reduce the impact of disturbances to operation devices. Protective devices such as relays are usually installed on various locations including power transmission lines, bus-bars, generators, transformers, and load nodes. Protective devices are equipped with sensors and act on the local level based on predefined settings that maintain the proper coordination between various relays \cite{7936473}. For instance, a primary protection relay trips and isolates a faulted transmission line. Also, some components such as turbine-governor units connected to electric generators require very detailed local protection schemes to operate properly. On the other hand, the global protection scheme focuses on the overall performance of the system without involvement of the local protection. It aims to detect abnormal system behavior, develop corrective actions, and respond in a quick and automatic way to prevent the propagation of a small disturbance to larger-scale events.  
          
\subsubsection{Measurement Components}
Measurement devices are mainly responsible for observing the performance of power system components. Measurement devices can be classified into system (central) measurement and component (local) measurement devices.  In the local level, measurements are passed to local controllers via spark communication links. For instance, generator units require an independent and massive measurement layer to monitor and maintain their performance, which could be mechanical, electrical or even physical measurements such as vibration sensors, rotor speed sensors, and magnetic field sensors. Global measurements, on the other hand, assess the performance of the power system as a whole. The transmission of global measurements heavily depends on two-way high-bandwidth communication technologies in order to access the information from the power grid and its components. These measurements are utilized to detect the propagation of a specific event to other components. For example, a faulted generator can be detected by measuring the variations in its reactive power flow \cite{6032699}.

\subsection{Cyber Layer}
A cyber layer can be identified as the layer that utilizes information and communication technology (ICT) and computer-aided platforms to gather, assess, and control the operation of power systems. It might be composed of communication channels, computation and control platforms, and monitoring systems.  

\subsubsection{Communication Channels}
ICT is a vital connecting bond between measurements and various cyber layers. Interface devices such as RTUs provide a two-way function in CPPS which are: (1) to transfer measured data via the communication layer, and (2) to execute decision-making signals coming from the control layer. RTUs are installed in various locations to capture the observability of the system states \cite{7936473}. Methods of communication between several components vary according to: system level, system scale, security constraint, priority, and hardware installation \cite{6473865}. Both local and wide area network environments are accompanied with several communication protocols to provide the proper communication. High capacity fiber optic cables are being widely used to connect between substations and system control centers in the transmission level at high transfer speed \cite{vellaithurai2015real}.

\subsubsection{Computation and Control platforms}
This layer is responsible for providing the proper control actions based on various power system assessment tools. Generally, control centers receive the measurements from field devices and pass them to operational processes, a decision is made and transmitted to actuators that apply a state change in the field devices. Both local and global centers utilize supervisory control and SCADA systems to handle the various computation and control algorithms \cite{vellaithurai2015real, 7796463, 7932449, 7936473}. Various monitoring screens are integrated to provide real-time information of the system components and status.

Each part of the power system has its own control algorithms, variables, and tools. In generation, terminal voltage and output power are the essential primary control algorithms. On the local level, generators have two control schemes: automatic voltage regulator, and governor control, whereas on the wide-area level, automatic generation control is used \cite{7936473}. To ensure safe operation of power flow through transmission lines, two control algorithms are utilized in the transmission system: state estimation and voltage-ampere reactive compensation. Two main algorithms are used in the distribution level control namely load shedding control, and advanced metering infrastructure.

\subsection{Interactions and Intra-actions}\label{interaction}
CPPS dependencies are classified into inter-actions and intra-actions, where the former studies the dependencies between various CPPS layers and the latter focuses on dependencies within a specific layer of a CPPS model. The complex interconnectivity between CPPS layers and the deep integration of ICT across all layers create further challenges to identify inter- and intra-dependencies. This section provides a brief explanation of these dependencies within the suggested CPPS model. 

The suggested model takes into consideration previous classifications as follows. The model identifies direct and indirect correlations among layers and sublayers. For instance, an event taking place in the global communication layer might directly propagate into the physical layer, whereas a fault at local protection devices might not be directly reflected in the main computation layer. Both inter- and intra-dependencies have been characterized in the suggested model. For example, steady-state power flow studies, and transient stability studies are utilized to assess the performance of power components in the physical layer. Physical layer and decision layer are dynamically interactive through the global communication layer, whereas results of the computation layer are not directly reflected on the physical layer. The suggested model gives insights on the common cause, cascading and escalating impacts. A power cyber-attack taking place in any cyber layer, either local or global, might cascade into the physical layer.

\subsection{Evaluating the Suggested Model}\label{modelevaluation}
As previously mentioned, the CPPS evaluation criteria can be used to measure the degree of competence of the suggested CPPS model. First, the suggested model provides a high accuracy outcome due to high matching between the model and the real system model. The suggested model can be scaled up to a specific level where the computational limits are not violated. However, co-simulation approaches can be leveraged to overcome this drawback. Also, the suggested model fulfills the fidelity feature since it provides a more detailed CPPS reducing the degree of approximations between various layers. High level of application-compatibility and topological-suitability is maintained. Different power system topologies, i.e., meshed and radial, and communication topologies, i.e., ring, star, and meshed, can be modeled. Finally, the suggested model can adapt to dynamic studies with high degree levels. Various time scales can be used for analysis and assessment.

\section{Conclusion}\label{conclusion}
This paper has classified system layers based-on their common, coupled, and shared functions. Also, interactions between the classified layers were identified and characterized, all possible interactions were enumerated, and they have been clustered based on their impact on the system. Furthermore, based on the overall function of the layer and types of its components, intra-action within the layers were characterized. The strategies developed in this paper for comprehensive classification of system layers and characterization of their inter- and intra-actions contributes towards the goal of accurate and detailed modeling of state transition and failure and attack propagation in CPPS. The accurate and detailed modeling of state transition and failure and attack propagation in CPPS is a necessary step towards reliability and resilience analysis, evaluation, and enhancement of CPPSs.

\section*{Acknowledgement}
This work was supported by the U.S. National Science Foundation (NSF) under Grant NSF 1847578.

\bibliographystyle{elsarticle-num}
\bibliography{References.bib}

\end{document}